\documentclass{PoS}

\usepackage{graphicx}                              
\usepackage{amsmath,amsfonts}
\graphicspath{{./pics/}}                       

\usepackage{epsfig}
\usepackage{amssymb}
\usepackage{amsfonts}
\usepackage{bbm}

\newcommand{\preprintline}{\newline
\vskip -5.2cm
\rightline{\parbox{4cm}{\large\rm  HU-EP-10/54\\ DESY 10-141}}
\vspace{4.2cm}}

\title{Effects of a potential fourth fermion generation on the upper and lower Higgs boson mass bounds \preprintline}
 
\ShortTitle{Effects of a fourth fermion generation on the Higgs boson mass bounds}
 
\author{\speaker{Philipp Gerhold}\\%
        Institut fur Physik, Humboldt Universit\"at zu Berlin, 12489 Berlin, Germany\\
        E-mail: \email{Philipp.Gerhold@physik.hu-berlin.de}}

\author{Karl Jansen\\
        NIC, DESY, 15738 Zeuthen, Germany\\
        E-mail: \email{Karl.Jansen@desy.de}}

\author{Jim Kallarackal\\
        Institut f\"ur Physik, Humboldt-Universit\"at zu Berlin, 12489 Berlin, Germany\\
        E-mail: \email{Jim.Kallarackal@physik.hu-berlin.de}}

\newcommand{\vs}{\vspace}
\newcommand{\hs}{\hspace}

\newcommand{\bdm}{\begin{displaymath}}
\newcommand{\edm}{\end{displaymath}}
\newcommand{\beq}{\begin{equation}}
\newcommand{\eeq}{\end{equation}}
\newcommand{\bea}{\begin{eqnarray}}
\newcommand{\eea}{\end{eqnarray}}
\newcommand{\bit}{\begin{itemize}}
\newcommand{\eit}{\end{itemize}}
\newcommand{\bc}{\begin{center}}
\newcommand{\ec}{\end{center}}
\newcommand{\re}{\relax{\rm I\kern-.18em R}}

\newcommand{\ie}{{\it i.e. }}

\newcommand{\latVol}[2]{\ifnum #1=#2 $#1^4$ \else $#1^3\times #2$\fi}
\newcommand{\lattice}[2]{\ifnum #1=#2 $#1^4$-lattice \else $#1^3\times #2$-lattice\fi}
\newcommand{\latticeX}[3]{\ifnum #1=#2 $#1^4$-lattice#3 \else $#1^3\times #2$-lattice#3\fi}
\newcommand{\lattices}[2]{\ifnum #1=#2 $#1^4$-lattice \else $#1^3\times #2$-lattices\fi}
\newcommand{\eq}[1]{Eq.~(\ref{#1})}

\newcommand{\fig}[1]{Fig.~\ref{#1}}

\newcommand{\Ref}[1]{Ref.~\cite{#1}}

\newcommand{\GEV}[1]{#1\,\mbox{GeV}}

\newcommand{\dslash}{\ensuremath\partial\kern-0.53em/}

\newcommand{\includeFigDouble}[7]{
\vs{-#6mm}
\bc
\begin{figure}[htb]
\centering
\begin{tabular}{cc}
\includegraphics[width=0.48\textwidth]{#1}
&
\includegraphics[width=0.48\textwidth]{#2}
\\
\hs{4mm}(a) & \hs{8mm}(b)  \\
\end{tabular}
\caption[#5]{#4}
\label{#3}
\vs{-2mm}
\end{figure}
\ec
\vs{-6mm}
\vs{-#7mm}
}

\abstract{We study the effect of a potential fourth fermion generation on the upper and lower Higgs boson mass bounds. 
This investigation is based on the numerical evaluation of a chirally invariant lattice Higgs-Yukawa model emulating 
the same Higgs-fermion coupling structure as in the Higgs sector of the electroweak Standard Model. In particular, 
the considered model obeys a Ginsparg-Wilson version of the underlying $\mbox{SU}(2)_L\times \mbox{U}(1)_Y$ symmetry, 
being a global symmetry here due to the neglection of gauge fields in this model. We present our results 
on the modification of the upper and lower Higgs boson mass bounds induced by the presence of a hypothetical very heavy 
fourth quark doublet. Finally, we compare these findings to the standard scenario of three fermion generations.}

\FullConference{35th International Conference of High Energy Physics\\
		 July 22-28, 2010\\
		 Paris, France}
 
\begin{document}

\section{Introduction}
\label{sec:Introduction}
 
The Sakharov explanation for the matter anti-matter asymmetry of the universe suffers from the CP-violating phase 
of the Standard Model (SM3) falling short by at least 10 orders of magnitude. In addition to this concern the Sakharov 
picture demands a first order electroweak phase transition, which is also objected in the framework of the SM3. However,
both of the above caveats might be addressable~\cite{Holdom:2009rf,Carena:2004ha} by the inclusion of a new fourth 
fermion generation into an extended version of the Standard Model (SM4). Despite the arguments against the existence
of a fourth fermion generation such a scenario nevertheless remains attractive for two reasons. Firstly, there
is a strong conceptual interest, since a new fermion generation would need to be very heavy, leading to rather large 
Yukawa coupling constants and thus to potentially strong interactions with the scalar sector of the theory. Secondly,
it has been argued~\cite{Holdom:2009rf} (and the references therein) that the fourth fermion generation is actually
{\it not} excluded by electroweak precision measurements, thus leaving the potential existence of a new fermion generation
an open question.

In our contribution, however, we do not present any statement arguing in favour or disfavour of a new fermion generation. Here, 
we simply assume its existence and focus on the arising consequences on the Higgs boson mass spectrum. With the advent
of the LHC this question will become of great phenomenological interest, since the Higgs boson mass bounds, in particular
the lower bound, depend significantly on the heaviest fermion mass. Demonstrating this effect will be the main
objective of the present work. 

Due to the large Yukawa coupling constants of the fourth fermion generation a non-perturbative computation is highly 
desirable. For this purpose we employ a lattice approach to investigate the strong Higgs-fermion interaction.
In fact, we follow here the same lattice strategy that has already been used in \Ref{Gerhold:JOINT} 
for the non-perturbative determination of the upper and lower Higgs boson mass bounds in the SM3. This latter approach
has the great advantage over the preceding lattice studies of Higgs-Yukawa models that it is the first being based on 
a consistent formulation of an exact lattice chiral symmetry~\cite{Luscher:1998pq}, which allows to emulate the chiral 
character of the Higgs-fermion coupling structure of the Standard Model on the lattice
in a conceptually fully controlled manner.

\section{Numerical Results}
\label{sec:model}

In order to evaluate the Higgs boson mass bounds we have implemented a lattice 
model of the pure Higgs-fermion sector of the Standard Model. To be more precise,
the Lagrangian of the targeted Euclidean continuum model we have in mind is given as
\begin{align}
\label{eq:StandardModelYuakwaCouplingStructure}
L_{HY} &= \bar t' \dslash t' + \bar b' \dslash b' + 
\frac{1}{2}\partial_\mu\varphi^{\dagger} \partial_\mu\varphi
+ \frac{1}{2}m_0^2\varphi^{\dagger}\varphi + \lambda\left(\varphi^{\dagger}\varphi\right)^2 
+ y_{b'} \left(\bar t', \bar b' \right)_L \varphi b'_R + y_{t'} \left(\bar t', \bar b' \right)_L \tilde\varphi t'_R  \notag\\
& + \mbox{c.c. of Yukawa interactions,} 
\end{align}
where we have restricted ourselves to the consideration of the heaviest quark
doublet, \ie the fourth generation doublet, which is labeled here $(t',b')$.
This restriction is reasonable, since the dynamics of the complex scalar doublet 
$\varphi$ ($\tilde \varphi = i\tau_2\varphi^*,\, \tau_i:\, \mbox{Pauli-matrices}$)
is dominated by the coupling to the heaviest fermions. For the same reason we also 
neglect any gauge fields in this approach. The quark fields nevertheless have
a colour index which actually leads to $N_c=3$ identical copies of the fermion doublet 
appearing in the model. However, for a first exploratory study of the fermionic influence
on the Higgs boson mass bounds we have set $N_c$ to 1 for simplicity.

The actual lattice implementation of the continuum model in \eq{eq:StandardModelYuakwaCouplingStructure}
has been discussed in detail in \Ref{Gerhold:JOINT}. Since the Yukawa interaction has a 
chiral structure, it is important to establish chiral symmetry also in the lattice approach. 
Here, it is only remarked that this has been a long-standing obstacle, which was finally found
to be circumventable by constructing the lattice equivalent of $\dslash$ as well as the left- and 
right-handed components of the quark fields $t'_{L,R}$, $b'_{L,R}$ on the basis of the Neuberger 
overlap operator~\cite{Luscher:1998pq, Neuberger:1998wv}. Following the proposition in \Ref{Luscher:1998pq} 
we have constructed a Higgs-Yukawa model with a global $SU(2)_L \times U(1)_Y$ symmetry on a 
$L_s^3\times L_t$-lattice.

Due to the triviality of the Higgs sector the targeted Higgs boson mass bounds actually depend on the 
non-removable, intrinsic cutoff parameter $\Lambda$ of the considered Higgs-Yukawa theory, which can be
defined as the inverse lattice spacing, \ie $\Lambda=1/a$. To determine these cutoff dependent bounds
for a given phenomenological scenario, \ie for given hypothetical masses of the fourth fermion generation,
the strategy is to evaluate the maximal interval of Higgs boson masses attainable within the framework of 
the considered Higgs-Yukawa model being in consistency with this phenomenological setup. The free parameters 
of the model, being the bare scalar mass $m_0$, the bare quartic coupling constant $\lambda$, and the Yukawa 
coupling constants $y_{t',b'}$ thus have to be tuned accordingly. The idea is to use the phenomenological
knowledge of the renormalized vacuum expectation value $v_r/a = \GEV{246}$ of the scalar field $\varphi$ as 
well as the hypothetical fourth generation quark masses $m_{t',b'}$ to fix the bare model parameters for a 
given cutoff $\Lambda$.

In lack of an additional matching condition a one-dimensional freedom is left open here, which can be parametrized
in terms of the quartic coupling constant $\lambda$. This freedom finally leads to the emergence of upper and lower
bounds on the Higgs boson mass. As expected from perturbation theory, one also finds numerically~\cite{Gerhold:JOINT} 
that the lightest and heaviest Higgs boson masses are obtained at vanishing and infinite bare quartic coupling constant, 
respectively. The lower mass bound will therefore be obtained at $\lambda=0$, while $\lambda=\infty$ will be adjusted to 
derive the upper bound.

Concerning the hypothetical masses of the fourth fermion generation quarks, we target here a mass degenerate scenario 
with $m_{t'}/a=m_{b'}/a=\GEV{700}$, which is somewhat above its tree-level unitarity upper bound~\cite{Chanowitz:1978mv}.
However, we are currently also investigating a set of other mass settings to study in particular the quark mass dependence of
the Higgs boson mass bounds.

For the eventual determination of the cutoff dependent Higgs boson mass bounds several series of 
Monte-Carlo calculations have been performed at different values of $\Lambda$ and on different lattice volumes. 
In order to tame finite volume effects as well as cutoff effects to an acceptable level, we have demanded as 
a minimal requirement that all particle masses $\hat m=m_{H}, m_{t'}, m_{b'}$ in lattice units fulfill $\hat m < 0.5$ 
and $\hat m\cdot L_{s,t}>3.5$, at least on the largest investigated lattice volumes. Assuming the Higgs boson mass $m_H$ 
to be around $\GEV{500-750}$ this allows to reach cutoff scales between $\GEV{1500}$ and $\GEV{3500}$ on a \latticeX{24}{32}{.} 
However, despite this setting strong finite volume effects are nevertheless expected induced by the massless Goldstone modes. 
It is known that these finite size effects are proportional to $1/L_s^2$ at leading order. An infinite volume extrapolation 
of the lattice data is therefore mandatory. Thanks to the multitude of investigated lattice volumes reaching from
$12^3\times 32$ to \lattices{24}{32} here, such an extrapolation could reliably be performed by fitting the finite 
volume data to the aforementioned leading order behaviour.

The obtained infinite volume results are finally presented in \fig{fig:PhysicalHiggsMassBounds}b. The numerical data for the
upper mass bound have moreover been fitted with the analytically expected functional form of the cutoff dependence of the upper
Higgs boson mass bound derived in \Ref{Luscher:1988uq}. It is given as
\bea
\label{eq:StrongCouplingLambdaScalingBeaviourMass}
\frac{m^{up}_{H}}{a} &=& A_m \cdot \left[\log(\Lambda^2/\mu^2) + B_m \right]^{-1/2}, 
\eea
with $A_m$, $B_m$ denoting the free fit parameters and $\mu$ being an arbitrary scale here. One learns from this presentation 
that the obtained results are indeed in good agreement with the expected logarithmic decline of the upper Higgs boson mass bound 
with increasing cutoff parameter $\Lambda$.

The reader may want to compare these findings to the upper and lower Higgs boson mass bounds previously derived in the SM3. The lattice 
results corresponding to that setup have been determined in \Ref{Gerhold:JOINT} and are summarized in \fig{fig:PhysicalHiggsMassBounds}a.
The main finding is that especially the lower bound is drastically shifted towards larger values in the presence of the assumed mass-degenerate 
fourth quark doublet. From this analysis it can be concluded that the usually expected light Higgs boson seems to be 
incompatible with a very heavy fourth fermion generation.

\includeFigDouble{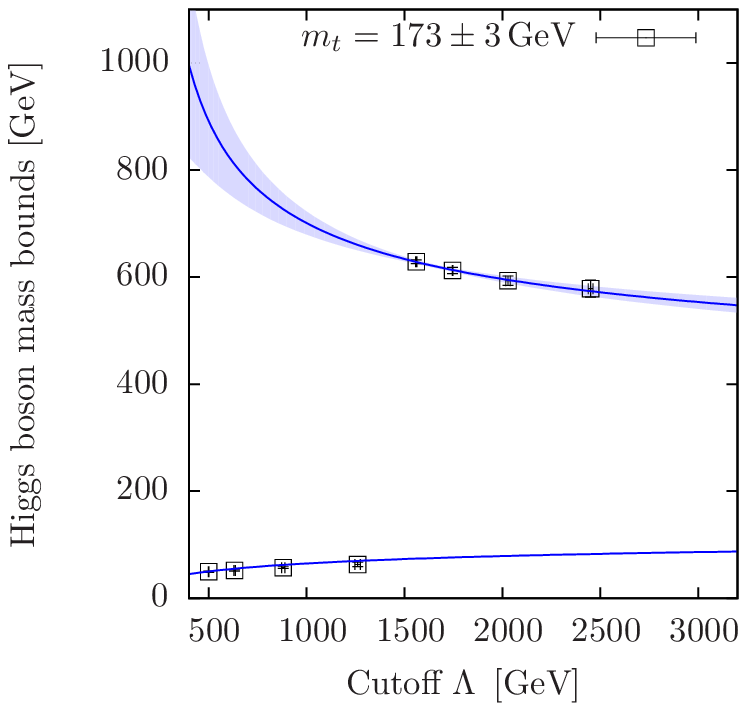}{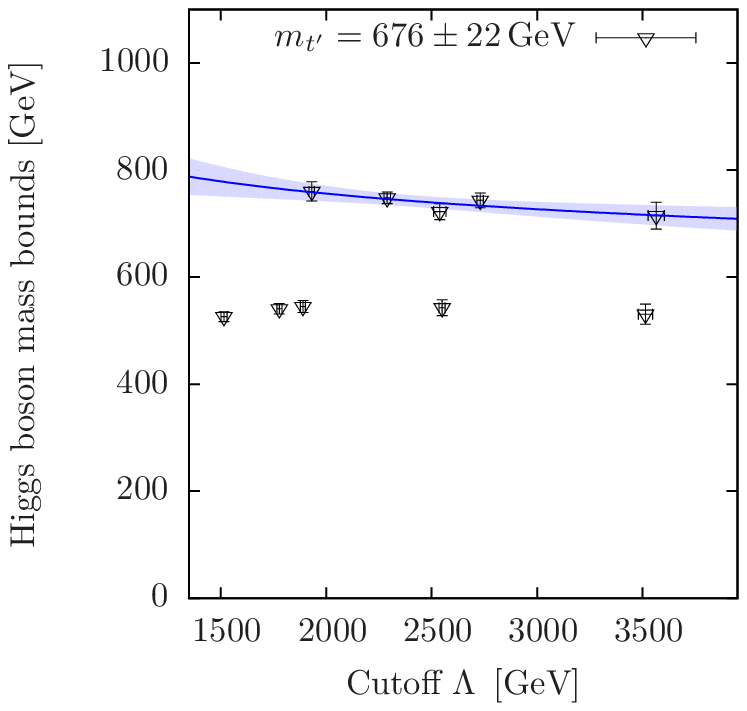}
{fig:PhysicalHiggsMassBounds}
{Upper and lower Higgs boson mass bounds are shown for $N_c=1$, $m_t=m_b=\GEV{173\pm 3}$ (a) and $N_c=1$, $m_{t'}=m_{b'}=\GEV{676\pm 22}$ (b).
Both upper bounds are each fitted with \eq{eq:StrongCouplingLambdaScalingBeaviourMass}. The lower bound in (a)
is also compared to a direct analytical computation depicted by the solid line as discussed in \Ref{Gerhold:JOINT}.
}
{InfiniteVolumeExtrapolationBothBounds}{3}{1}

\bibliographystyle{unsrtOwnNoTitles}
\bibliography{Proceedings}
  
\end{document}